# Electron-optical phonon scattering in quantum wells in a tilted quantizing magnetic field

M.P. Telenkov[*], Yu.A. Mityagin, D.S.Korchagin

P.N. Lebedev Physical Institute of the Russian Academy of Sciences, 119991, Moscow, Russia.

Keywords: quantum wells, tilted quantizing magnetic field, optical phonon scattering

The electron scattering with longitudinal polar optical phonons in a quantizing magnetic field tilted to the plane of quantum well layers is studied. The scattering rate's behavior at variation of the magnetic field magnitude and orientation is established.

* telenkovmp@lebedev.ru

# 1. Introduction

The interaction of charge carriers with optical vibrations of the crystal lattice plays an important role in semiconductor physics. In particular, electron scattering processes with optical phonon emission are one of the main mechanisms of charge carrier energy relaxation in quantum well structures [1-3]. For these reasons, electron-optical phonon scattering in quantum wells has been extensively studied [4-6], including that in a quantizing magnetic field [7-9].

However, the scattering processes were studied in a quantizing magnetic field perpendicular to the layers of the quantum well.

At the same time, there are indications [10, 11] that by tilting the magnetic field relative to the plane of the quantum well structure, one can significantly affect the various transitions between the Landau levels.

This paper presents formulas for the electron scattering rate due to longitudinal polar optical phonon emission in a quantizing magnetic field that is tilted relative to the quantum well layers. The study explores how the electron-phonon scattering processes are effected by variations in both the magnitude and orientation of the magnetic field.

# 2. Spectrum of an electron in a quantum well in a tilted quantizing magnetic field

The magnetic field effect on the electron spectrum is largely determined by its orientation relative to the growth axis of the quantum well (the plane of its layers).

When the magnetic field is aligned parallel to the layers of the quantum well (perpendicular to the z-axis of its growth), it does not alter the energy spectrum's structure. The energy spectrum remains composed of a series of continuous two-dimensional subbands. The effect of a moderate magnetic field in this scenario mainly results in a shift of each subband by the value of

$$\Delta_v \left( B_\parallel \right) = \frac{e^2}{2 m_w c^2} \left( \delta z \right)_v^2 \cdot B_\parallel^2, \qquad (1)$$

proportional to both the square of the magnetic field strength $B_\parallel$ and the square of the mean-square fluctuation $\left( \delta z \right)_v$ of the electron coordinate along the growth axis of the quantum well [12]. When the cyclotron energy essentially exceeds the distance between subbands at sufficiently large magnetic fields, the magnetic field changes the subband dispersion law [13]. With increasing magnetic field, the dispersion law of the subband gradually flattens along the momentum perpendicular to the magnetic field and the quantum well growth axis, and only at very high values of magnetic fields (the magnetic length is much smaller than the width of the quantum well), the electron spectrum transforms to the one-dimensional Landau subbands corresponding to the bulk material of the quantum well [14].

The situation changes substantially when the magnetic field is applied perpendicularly to the layers of the quantum well (along the z-axis of its growth). In this case, the magnetic field changes the character of the energy spectrum - it transforms each two-dimensional subband of quantum well into a discrete set of Landau levels [15]

$$E_{(v,n)} = \varepsilon_v + \hbar\omega_c \left(n + \frac{1}{2}\right), \qquad (2)$$

where $\varepsilon_v$ - subband energy, $\omega_c = \dfrac{eB}{mc}$ - cyclotron frequency, n=0,1,2,3,… - Landau level number. The Landau level degeneracy

$$\alpha = \frac{1}{\pi \ell^2} = \frac{e}{\pi \hbar c} \cdot B = 4.9 \cdot 10^{10} \cdot B_\perp \; \frac{cm^{-2}}{T} \qquad (3)$$

grows linearly with the magnetic field strength and reaches macroscopic values in quantizing magnetic fields.

In the case when, in addition to the quantizing magnetic field $\mathbf{B}_\perp = B_\perp \mathbf{e}_z$, a magnetic field $\mathbf{B}_\parallel = B_\parallel \mathbf{e}_y$ parallel to the layers of the quantum well is applied (i.e., the total magnetic field $\mathbf{B} = B_\perp \mathbf{e}_z + B_\parallel \mathbf{e}_y$ is tilted to the plane of the quantum well layers), the Hamiltonian for electron envelope wave-function [16]

$$\hat{H} = \left(\hat{\mathbf{p}} + \frac{e}{c}\mathbf{A}\right) \frac{1}{2m(z)} \left(\hat{\mathbf{p}} + \frac{e}{c}\mathbf{A}\right) + U(z) \qquad (4)$$

takes in the Landau gauge $\mathbf{A} = (B_\parallel z - B_\perp y)\mathbf{e}_x$, the form

$$\hat{H} = \hat{\mathbf{p}} \frac{1}{2m(z)} \hat{\mathbf{p}} + U(z) + \frac{m_w}{m(z)} \left[ \left(\omega_\parallel z - \omega_\perp y\right) \hat{p}_x + \frac{m_w}{2} \left(\omega_\parallel z - \omega_\perp y\right)^2 \right]. \qquad (5)$$

Here $U(z)$ is the potential profile of the quantum well, $m(z)$ is the electron effective mass ($m_w$ in the well and $m_b$ in the barrier), $\omega_\perp = \dfrac{eB_\perp}{m_w c}$ and $\omega_\parallel = \dfrac{eB_\parallel}{m_w c}$ are the cyclotron frequencies for the magnetic field component along the growth axis of the structure ($B_\perp$) and the magnetic field component in the plane of its layers ($B_\parallel$). respectively.

Since $\widehat{H}\widehat{p}_x - \widehat{p}_x\widehat{H} = 0$ then, we can construct a basis of stationary states with a certain momentum projection value on the x-axis. The wave functions of such a basis have the form

$$\Psi(x,y,z) = \frac{\exp(ik_x x)}{\sqrt{L}} \psi\left(y - k_x \ell_\perp^2, z\right), \qquad (6)$$

where $\ell_\perp = \sqrt{\frac{\hbar}{m\omega_\perp}} = \sqrt{\frac{\hbar c}{eB_\perp}}$ is the magnetic length for the transverse component of the magnetic field. The electron energy levels and wave functions of stationary states are defined by the two-dimensional Hamiltonian [17]

$$\widehat{H}_{2D} = \widehat{H}_\perp + \widehat{H}_\parallel \qquad (7)$$

where

$$\widehat{H}_\perp = -\frac{\partial}{\partial z}\frac{\hbar^2}{2m(z)}\frac{\partial}{\partial z} + U(z) + \frac{m_w}{m(z)}\left[\frac{\widehat{p}_y^2}{2m_w} + \frac{m_w \omega_\perp^2}{2} y^2\right] \qquad (8)$$

is the electron Hamiltonian when the magnetic field is directed along the growth axis of the structure. The term

$$\widehat{H}_\parallel = +\frac{m_w}{m(z)}\frac{m_w \omega_\parallel^2 z^2}{2} - \frac{m_w}{m(z)} m_w \omega_\parallel \omega_\perp zy \qquad (9)$$

is due to the magnetic field component being parallel to the layers of the quantum well.

The variables in the Schrödinger equation with the Hamiltonian (8) are separable. The energy levels have the form (2), and the wave functions are given by the expression [16]

$$\Psi(x,y,z) = \frac{\exp(ik_x x)}{\sqrt{L}} \varphi_\nu(z)\Phi_n(y), \qquad (10)$$

where $\varphi_\nu(z)$ is the wave function of the subband energy level (eigen wave-function of the Hamiltonian $\widehat{H}_z = -\frac{\partial}{\partial z}\frac{\hbar^2}{2m(z)}\frac{\partial}{\partial z} + U(z)$), $\Phi_n(y)$ - the wave function of the nth (n=0,1,2,...) energy level of a linear harmonic oscillator with cyclotron frequency $\omega_\perp$, L is the sample size in the structure layer.

Note that the relation $m_w/m(z)$ in the third term of the Hamiltonian (7) effectively lowers the barrier height with increasing the Landau level number n by adding to $\hat{H}_z$ the term $-\left(1-\frac{m_w}{m(z)}\right)\hbar\omega_\perp\left(n+\frac{1}{2}\right)$. This leads to a dependence on n of the subband energy levels $\varepsilon$ and their wave functions $\varphi(z)$ [15]. This effect is similar in nature to dimensionality transformation at large wave-vector values in quantum wells [18]. However, it is known that the dependence of the stationary states in a quantum well on the barrier height is significant only for levels close to the continuous spectrum. Therefore, we will further neglect this effect due to its smallness for the lower subbands in the considered sufficiently deep (several hundreds of meV) and wide (more than 5 nm) quantum wells at reasonable values of Landau level numbers.

The matrix of the Hamiltonian (5) in the basis of wave functions (6) is diagonal by $k_x$, and the matrix element at $k_{x1} = k_{x2}$

$$\left\langle \frac{\exp(ik_x x)}{\sqrt{L}}\psi_1(y-k_x\ell_\perp^2,z)\left|\hat{H}\right|\frac{\exp(ik_x x)}{\sqrt{L}}\psi_2(y-k_x\ell_\perp^2,z)\right\rangle = \\ = \left\langle \psi_1(y,z)\left|\hat{H}_{2D}\right|\psi_2(y,z)\right\rangle \qquad (11)$$

does not depend on $k_x$. Therefore, in the tilted quantizing magnetic field, the Landau level degeneracy is determined only by the quantizing component $B_\perp$ of the magnetic field and is given by the expression (3).

The matrix element between the Landau level $(v_1,n_1)$ and $(v_2,n_2)$ is given by the expression [19]

$$\left\langle v_1,n_1\left|\hat{H}_{2D}\right|v_2,n_2\right\rangle = \left[\varepsilon_{v_1}+\hbar\omega_\perp(n+1/2)\right]\delta_{v_1,v_2}\delta_{n_1,n_2} + \\ + \frac{m_w\omega_\parallel^2}{2}\langle z^2\rangle_{v_1,v_2}\delta_{n_1,n_2} - m_w\hbar\omega_\parallel\sqrt{\hbar\omega_\perp}\sqrt{\frac{m_w}{2\hbar^2}}\langle z\rangle_{v_1,v_2}\times \\ \times\left[\sqrt{n_2+1}\cdot\delta_{n_1,n_2+1}+\sqrt{n_2}\cdot\delta_{n_1,n_2-1}\right] \qquad (12)$$

In a quantum well in a tilted magnetic field, there are two scales of energy - cyclotron energy and intersubband spacing. We will be interested in the case when the cyclotron energy is several times smaller than the intersubband spacing. In this case, in matrix (12), we can neglect the coupling between subbands (elements with $v_1 \neq v_2$) and diagonalize it analytically [10,20-22]. As a result, the following expressions for the Landau levels and wave functions of stationary states are obtained

$$E_{(v,n)} = \varepsilon_v + \delta\varepsilon_v(B_\parallel) + \hbar\omega_\perp\left(n + \frac{1}{2}\right) \quad (13)$$

and

$$\Psi(x,y,z) = \frac{\exp(ik_x x)}{\sqrt{L}} \varphi_v(z) \Phi_n\left(y - k_x \ell_\perp^2 - \langle z \rangle_v tg\theta\right). \quad (14)$$

Here

$$\delta\varepsilon_v(B_\parallel) = \frac{m_w \omega_\parallel^2}{2}(\delta z)_v^2 = \frac{e^2}{2m_w c^2}(\delta z)_v^2 \cdot B_\parallel^2. \quad (15)$$

$$\langle z \rangle_v = \int dz\, \varphi^*(z) z \varphi(z) \quad (16)$$

is the mean value of the electron z-coordinate, $(\delta z)_v$ - its standard deviation, $\theta$ -- tilted magnetic field angle to the quantum well's growth axis ($tg\theta = \frac{B_\parallel}{B_\perp}$).

As can be seen, the Landau quantization is caused only by the magnetic field component $B_\perp$ perpendicular to the layers of the heterostructure.

The component $B_\parallel$ does not lead to the quantization of the electron energy. Its primary effect on the energy spectrum is to shift each subband by an amount that is proportional to $B_\parallel^2$ and $(\delta z)_v^2$ associated with the states of the given subband. The $B_\parallel$ also leads to the coupling of quantizing systems - the magnetic component $\Phi(y)$ of the wave-function becomes dependent on the subband wave-function $\varphi(z)$ – there is a shift of the center of the linear harmonic oscillator by a value $\langle z \rangle_v tg\theta$, proportional to $B_\parallel$.

We neglect the spin splitting of Landau levels because it is minimal within the class of structures and magnetic field range we are considering. Our focus is on quantum wells made of non-magnetic III-V semiconductors, where in the magnetic field range of 1 to 10 T, the Zeeman splitting is significantly smaller than the width of the Landau levels, due to the small Landé factor [7, 23, 24].

## 3. Electron-optical phonon scattering rate

To describe the kinetics of electrons in the Landau level system, it is essential to calculate the electron fluxes between levels. Specifically, these fluxes determine a system of rate equations [24-31]

$$\frac{dN_i}{dt} = \sum_f J_{f \to i} - \sum_f J_{i \to f}. \quad (17)$$

Here $N_i$ is the number of electrons at the Landau level $i = (\nu_i, n_i)$ per unit area of the quantum well (the population of the Landau level).

In the Fermi's rule approximation [32], the intensity of electron transitions (measured as the number of scattering events per unit time per unit area of the quantum well) from Landau level $i$ to Landau level $j$, accompanied by the emission of longitudinal polar optical phonons, is expressed as follows[7]:

$$J_{i \to f} = \frac{1}{\tau_{i \to f}} \cdot N_i \left[ 1 - \frac{N_f}{\alpha} \right], \quad (18)$$

where

$$\frac{1}{\tau_{i \to f}} = A_{LO} \binom{i}{f} \times \delta\left(E_i - E_f - \hbar\omega_{LO}\right) \quad (19)$$

is the electron-phonon scattering rate corresponding to this transition,

$$A_{LO}\binom{i}{f} = \frac{2}{\alpha L^2} \sum_{\mathbf{q}} \sum_{k_i, k_f} \frac{2\pi}{\hbar} \left| \langle \nu_f, n_f, k_f | \hat{H}_{e-ph}(\mathbf{q}) | \nu_i, n_i, k_i \rangle \right|^2 =$$

$$= \frac{L^3}{8\pi^4 \alpha \hbar} \int d\mathbf{q} \int dk_i \int dk_f \left| \langle \nu_f, n_f, k_f | \hat{H}_{e-ph}(\mathbf{q}) | \nu_i, n_i, k_i \rangle \right|^2, \quad (20)$$

is the amplitude of the transition.

$$\langle \nu_f, n_f, k_f | \hat{H}_{e-ph}(\mathbf{q}) | \nu_i, n_i, k_i \rangle = \int d\mathbf{r} \Psi^*_{(f,k_f)}(\mathbf{r}) \hat{H}_{e-ph}(\mathbf{q}) \Psi_{(i,k_i)}(\mathbf{r}), \quad (21)$$

$$\hat{H}_{e-ph}(\mathbf{q}) = i \left( 2\pi \hbar \omega_{LO} \frac{e^2}{\varepsilon_p L^3} \right)^{1/2} \frac{\exp(-i\mathbf{q}\mathbf{r})}{q} \left[ 1 + N_B(\hbar\omega(\mathbf{q})) \right]^{1/2}, \quad (22)$$

is the Fröhlich Hamiltonian for the interaction of electrons with longitudinal polar optical phonons (LO-phonons), $N_B(x) = \frac{1}{\exp(x/T_L) - 1}$ is the Planck distribution function, $T_L$ is lattice temperature, $\varepsilon_p^{-1} = \varepsilon_\infty^{-1} - \varepsilon_s^{-1}$, $\varepsilon_\infty$ is high-frequency dielectric permittivity, $\varepsilon_s$ is static dielectric permittivity. $\omega_{LO}$ is LO-phonon frequency. The Dirac's delta- function expresses the energy conservation law

$$E_i - E_f - \hbar\omega_{LO} = 0. \tag{23}$$

for the LO-phonon emission by an electron:

The amplitude of the scattering rate involves a multiple integral with a Coulomb singularity, which complicates its calculation and makes it time-consuming. By utilizing the wave functions (14), we were able to perform an analytical integration in (20), significantly reducing the complexity of the integral. As a result, we derived the following expression for the electron-phonon scattering amplitude:

$$A_{LO}\binom{i}{f} = \frac{1}{2\hbar}\hbar\omega_{LO}\frac{e^2}{\varepsilon_p \ell_\perp}\left[1 + N_B(\hbar\omega_{LO})\right]\frac{1}{n_f! n_i!} \times$$
$$\times \int_0^{+\infty} d\beta\, f^{(LO)}_{\nu_i,\nu_f}\left(\frac{\beta}{\ell_\perp}\right)\int_0^{2\pi} d\varphi \exp(-\tilde\beta^2/2) P^2_{n_i,n_f}\left(\frac{\tilde\beta^2}{2}\right), \tag{24}$$

where

$$f^{(LO)}_{\nu_i,\nu_f}(\gamma) = \int dz_1 dz_2\, \varphi^*_{\nu_f}(z_1)\varphi_{\nu_i}(z_1)\varphi_{\nu_f}(z_2)\varphi^*_{\nu_i}(z_2)\exp(-\gamma|z_2 - z_1|), \tag{25}$$

$$P_{n_i,n_f}(x) = \sum_{j=0}^{\min\{n_i,n_f\}} (-1)^j \binom{n_i}{j}\binom{n_f}{j} j!\, x^{\frac{(n_i+n_f-2j)}{2}}. \tag{26}$$

$$\tilde{\boldsymbol{\beta}} = \left(\beta\cos\varphi - \xi_{\nu_f,\nu_i}\right)\mathbf{e}_x + \beta\sin\varphi\,\mathbf{e}_y, \tag{27}$$

$$\xi_{\nu_f,\nu_i} = \sqrt{\frac{e}{\hbar c B_\perp}}\cdot\left[\langle z\rangle_{\nu_f} - \langle z\rangle_{\nu_i}\right]\cdot B_\|. \tag{28}$$

In the expression (24) for the scattering amplitude, the integral (25) solely contains information about the potential profile of the heterostructure, incorporating the subband wave functions $\varphi(z)$. Therefore, when considering the various transitions between two subbands or within a single subband in the given structure, it is unnecessary to calculate the function $f^{(LO)}_{\nu_i,\nu_f}(\gamma)$ for each transition individually. Instead, it is sufficient to compute this function only once and then use it to determine the scattering rates for transitions $(\nu_i, n_i) \to (\nu_f, n_f)$ with all possible $n_i$ and $n_f$ at any magnetic field strength. Essentially, we have simplified the problem to calculating the double integral of the smooth function.

Also note that at $B_\| = 0$ or $\langle z\rangle_{\nu_f} = \langle z\rangle_{\nu_i}$, the parameter $\xi_{\nu_f,\nu_i} = 0$. In this case, the expression for the transition amplitude becomes even simpler

$$A_{LO}\begin{pmatrix}i\\f\end{pmatrix} = \frac{\pi}{\hbar}\hbar\omega_{LO}\frac{e^2}{\varepsilon_p \ell_\perp}\left[1+N_B(\hbar\omega_{LO})\right]\frac{1}{n_f!n_i!}\times$$
$$\times\int_0^{+\infty} d\beta\, f^{(LO)}_{\nu_i,\nu_f}\left(\frac{\beta}{\ell_\perp}\right)\exp(-\beta^2/2)P^2_{n_i,n_f}\left(\frac{\beta^2}{2}\right)$$
, (29)

## 4. Inter-Landau level scattering with optical phonon emission

The primary effect of $B_\parallel$ on the electron-phonon scattering rate can be conditionally divided into two parts.

First, the magnetic field component parallel to the layers leads to a shift of the subbands by an amount (1) proportional to both $B_\parallel^2$ and $(\delta z)_\nu^2$ of the coordinate z in the states of a given subband. Accordingly, $B_\parallel$ enters the energy conservation law (23).

Secondly, the parallel component $B_\parallel$ causes the oscillator center to shift within the plane of the quantum well layers, with the shift being proportional to the average coordinate $\langle z\rangle_\nu$ along the growth axis of the structure. Consequently, the scattering rate amplitude can depend not only on the quantizing component $B_\perp$ but also on $B_\parallel$.

Furthermore, as can be directly observed from the obtained expression (24), $B_\parallel$ enters the transition amplitude $A_{LO}$ only via the parameter $\xi$ defined by expression (28). When this parameter is zero, the amplitude $A_{LO}$ does not depend on $B_\parallel$. At $B_\parallel \neq 0$, the parameter $\xi$ is zero if and only if the mean coordinates in the initial and final states are identical ($\langle z\rangle_{\nu_i} = \langle z\rangle_{\nu_f}$).

Obviously, this situation applies to all intra-subband scattering processes. For an intra-subband transition, $\nu_i = \nu_f$ and, therefore, $\xi = 0$. Accordingly, the amplitude of all transitions between Landau levels within a single subband is independent of $B_\parallel$. Since $B_\parallel$ does not affect the spacing between the Landau levels of a single subband, the resonance condition for intra-subband transitions is also unaffected by $B_\parallel$. Therefore, the magnetic field component parallel to the layers of the quantum well does not affect the intra-subband electron-phonon scattering processes.

The situation changes for inter-subband transitions. Following the oscillatory theorem, the wave function $\varphi_\nu(z)$ of the $\nu$-th subband has $\nu$ zeros. Therefore, the wave function of the higher subband behaves in a more complex manner, leading to an increase in the RMS-deviation $(\delta z)_\nu^2$ with rising $\nu$. Therefore, according to (15), the magnetic field component $B_\parallel$ enlarges the separation between the subbands. This relative shift of the subbands means that the energy conservation law (23) for scattering is satisfied at another value of the quantizing component $B_\perp$ of the magnetic field.

Considering (13), the energy conservation law (23) for the transition is expressed as follows:

$$\Delta\varepsilon_{if} + \delta\varepsilon_{if}(B_\parallel) - \hbar\omega_{LO} + \hbar\omega_\perp(n_i - n_f) = 0, \qquad (30)$$

where

$$\Delta\varepsilon_{\nu_i,\nu_f} = \varepsilon_{\nu_i} - \varepsilon_{\nu_f} \qquad (31)$$

is the distance between the subbands at $\mathbf{B}=0$ for transitions from the upper subband to the lower one,

$$\delta\varepsilon_{\nu_i,\nu_f}(B_\parallel) = \frac{e^2}{2m_w c^2}\left[(\delta z)_{\nu_i}^2 - (\delta z)_{\nu_f}^2\right]\cdot B_\parallel^2 \qquad (32)$$

is an increment of the inter-subband distance due to $B_\parallel$. For transitions from the upper subband to the lower subband, this increment of the inter-subband distance is positive and grows with increasing parallel component $B_\parallel$ (Fig. 1).

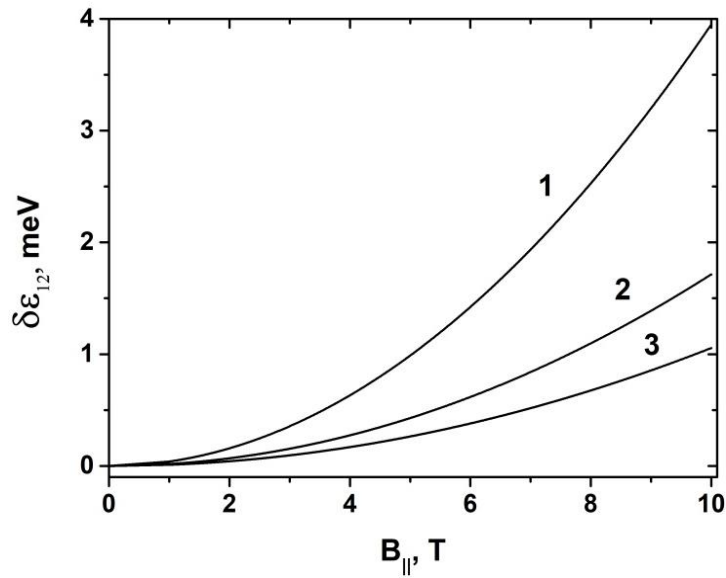

**Figure 1.** Dependence of the distance between two lower subbands of the quantum well on the magnetic field component parallel to its layers. Different curves correspond to different values of the width *a* of the quantum well: 1 - *a*=25 nm; 2 - *a*=15 nm; 3 - *a*=10 nm. The data are given for GaAs/Al$_{0.3}$Ga$_{0.7}$As quantum well.

From (30) we find

$$B_\perp = B_\perp^{(0)} + \delta B_\perp, \qquad (33)$$

where

$$B_\perp^{(0)} = \frac{mc}{e\hbar} \frac{(\Delta\varepsilon_{if} - \hbar\omega_{LO})}{\Delta n} \qquad (34)$$

is the value of the magnetic field at which the transition resonance occurs at $B_\parallel = 0$,

$$\delta B_\perp = \frac{mc}{e\hbar} \frac{\delta\varepsilon_{if}(B_\parallel)}{\Delta n} \qquad (35)$$

is the shift of the resonance position at switching on $B_\parallel$, $\Delta n = n_f - n_i$ - the change of the Landau level number of the electron at scattering. Accordingly, for the relative magnitude of the shift of the resonance, we obtain

$$\frac{\delta B_\perp}{B_\perp^{(0)}} = \frac{\delta\varepsilon_{if}(B_\parallel)}{\Delta\varepsilon_{if}} \frac{1}{1 - \frac{\hbar\omega_{LO}}{\Delta\varepsilon_{if}}}. \qquad (36)$$

As can be seen from (36), the absolute value of the shift is proportional to the increment of the inter-subband distance $\delta\varepsilon_{if}(B_\parallel)$. The magnitude $\delta\varepsilon_{if}(B_\parallel)$ increases with increasing width of the quantum well (Fig. 1). The increase is due to the fact that with the growth of the width of the quantum well, the area of localization of the wave function increases, and, accordingly, the standard deviation of each of the subbands increases. In order to estimate the type of this dependence, we consider deep levels in a rectangular quantum well with width *a*, in which the overwhelming part of the wave function is located in the well ( $\frac{\hbar}{\sqrt{2m(U_0 - \varepsilon)}} \ll a$ ). In this case, we can neglect the penetration of the wave function into the barrier and make an estimate for an infinitely deep quantum well. In this case, we obtain

$$(\delta z)_v^2 = \frac{a^2}{12}\left(1 - \frac{6}{(\pi v)^2}\right), \qquad (37)$$

Correspondingly,

$$\delta\varepsilon_{if}\left(B_{\parallel}\right) = \frac{e^2}{4\pi^2 mc^2}\left[\frac{1}{v_f^2} - \frac{1}{v_i^2}\right] \cdot a^2 \cdot B_{\parallel}^2 \tag{38}$$

Thus, we see that the magnitude of the transition resonance shift depends essentially on the width of the quantum well $a$ - it increases with increasing width approximately as $a^2$.

From expression (36) we can directly see that the ratio $\hbar\omega_{LO}/\Delta\varepsilon_{if}$ determines the sign of the shift of the resonance position. Since for the transitions from the upper subband to the lower one, $\delta\varepsilon_{if}(B_{\parallel}) > 0$ and $\Delta\varepsilon_{if} > 0$, the presence of the multiplier $\frac{1}{1-\hbar\omega_{LO}/\Delta\varepsilon_{if}}$ in (36) leads to the fact that the sign of the shift depends on the ratio of the optical phonon energy $\hbar\omega_{LO}$ to the distance between the subbands $\Delta\varepsilon_{if}$ at $B_{\parallel} = 0$. In the case when $\Delta\varepsilon_{if} > \hbar\omega_{LO}$, $\delta B_{\perp} = B_{\perp} - B_{\perp}^{(0)} > 0$, i.e., the shift of resonances of all transitions occurs towards higher values of the quantizing component $B_{\perp}$ of the magnetic field. In the case when $\Delta\varepsilon_{if} < \hbar\omega_{LO}$, $\delta B_{\perp} = B_{\perp} - B_{\perp}^{(0)} < 0$, i.e., the shift of resonances of all transitions occurs towards smaller values of $B_{\perp}$. In addition, as can be seen, a special case is represented by the situation when $\Delta\varepsilon_{if} = \hbar\omega_{LO}$. Therefore, let us consider these three cases separately.

1). The case $\Delta\varepsilon_{if} > \hbar\omega_{LO}$ (the upper subband lays above the optical phonon energy) is considered in [7] in some detail for $B_{\parallel} = 0$.

In this case, the energy conservation law in scattering is satisfied for a discrete set of magnetic field values determined by expression (34). Therefore, the dependence of the rate of a single transition on the magnetic field has a resonance character. As a result, the magnetic field dependence of the total scattering rate from a given Landau level to the lower subband $\frac{1}{\tau^{(tot)}_{(v_i,n_i)}} = \sum_{n_f=n_i+1}^{+\infty} \frac{1}{\tau_{(v_i,n_i)\to(v_f,n_f)}}$ has an oscillatory character, which comes to a monotonic drop at $B > B_{\perp,max}^{(0)}$. (Fig. 2).

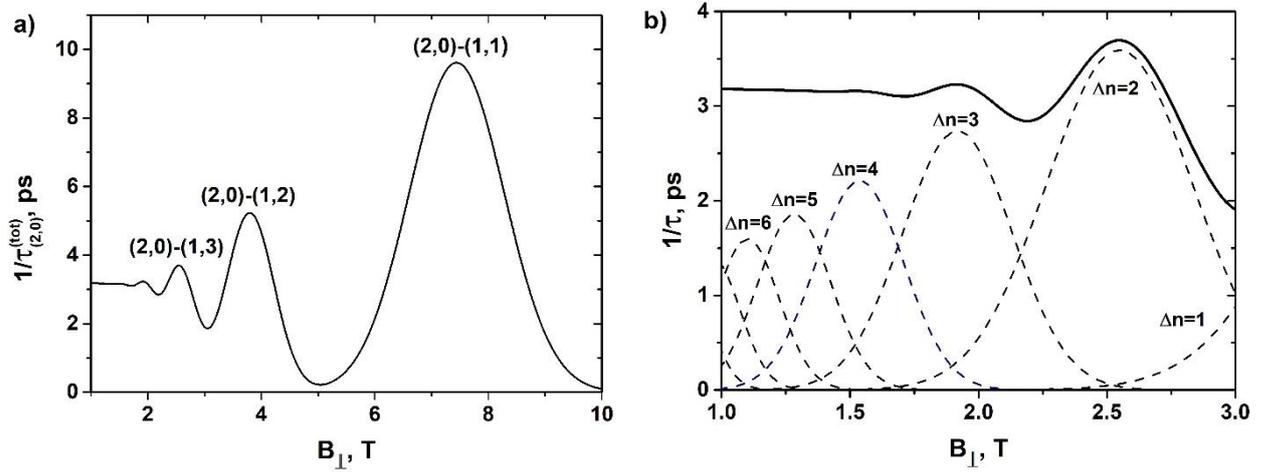

**Figure 2.** Magnetic field dependence of the total transition rate of an electron from the (2,0) level due to the optical phonon emission in a quantum well with an inter-subband distance exceeding the optical phonon energy. The dotted line indicates the rates of transitions from the level (2,0) to individual Landau levels (1,Δn) of the lower subband. The data are given for a GaAs/Al$_{0.3}$Ga$_{0.7}$As quantum well of 15 nm width ( $\Delta\varepsilon_{21} \approx 49$ meV $> \hbar\omega_{LO} \approx 36$ meV ).

The broadening of the Landau levels leads to the replacement of the Dirac delta function in the expression (19) by a form-factor of finite width: The type of form-factor and its width depend on the proportion between collision broadening (due to impurities, electron-electron interaction, interaction with the lattice, etc.) and inhomogeneous broadening due to fluctuations in the quantum well potential profile [7], the ratio between collision broadening and temperature [33-35], and the ratio between the radius of action of scatters and magnetic length (in particular, the electron density and Landau level degeneracy) [36,37]. However, in any case, the form-factor has a maximum when the energy conservation law (23) is fulfilled. Since the effect under discussion is mainly related to changes in the energy conservation law due to $B_\parallel$, the specific type of form-factor and its width do not play an underlying role for the effect. Therefore, we replace the Dirac delta function in the expression (19) with a Gaussian

$$F_{LO}\left(E_i - E_f - \hbar\omega_{LO}\right) = \frac{1}{\sqrt{2\pi}\left(\sqrt{2}\Gamma\right)} \exp\left(-\frac{\left(E_i - E_f - \hbar\omega_{LO}\right)^2}{2\left(\sqrt{2}\Gamma\right)^2}\right), \qquad (39)$$

with a typical broadening Γ=1 meV.

As can be seen from (34), resonances occur only for transitions with $\Delta n = n_f - n_i > 0$. Therefore, the discrete set (34) is limited from above by the magnetic field value $B_{\perp,\max}^{(0)} = \frac{mc}{\hbar e} \cdot (\Delta \varepsilon_{if} - \hbar \omega_{LO})$, which corresponds to $\Delta n = 1$. At the same time, there is no restriction from below - $\Delta n$ can be as large as you like, and, consequently, the resonance field can be correspondingly low.

Resonances of transitions with a larger change of the Landau level number $\Delta n$ and, correspondingly, with a larger difference in the wave functions of the initial and final states (in particular, $\Delta n$ is the difference in the number of zeros of the wave functions of the initial and final states) take place at lower magnetic fields. Therefore, the peak amplitude at the maxima of the total scattering rate decreases as the magnetic field decreases. The distance between the resonances of neighboring transitions and the difference in the magnitude of the maxima decrease with decreasing magnetic field. As a result, at relatively small magnetic fields, the dependence smooths out (becomes weakly oscillatory) due to the summation of closely spaced peaks with close amplitudes.

The component $B_\parallel$ shifts the resonance of each transition towards higher values of $B_\perp$, following (35) (Fig. 3).

The effect of $B_\parallel$ on the transition rate amplitude $A_{LO}$ is determined by the symmetry of the potential profile of the quantum well. In a symmetric quantum well, a wave function of a subband is either even or odd

$$\varphi_\nu(-z) = (-1)^\nu \varphi_\nu(z) \qquad (40)$$

Therefore, $\langle z \rangle_{\nu_i} = \langle z \rangle_{\nu_f}$. Accordingly, $\xi_{\nu_i,\nu_f} = 0$, and the amplitude of the scattering rate $A_{LO}$ does not depend on $B_\parallel$. However, the amplitude increases significantly as the quantizing component $B_\perp$ of the magnetic field rises (Fig. 4). As a result, the $B_\parallel$-induced shift of the resonance towards larger values of $B_\perp$ is accompanied by the increase of the scattering rate $1/\tau_{i \to f}$ at the resonance.

These shifts of the resonance peaks toward higher values of the quantizing component $B_\perp$ of the magnetic field, along with the associated increase in their amplitudes, are evident in the dependence of the total scattering rate $1/\tau_i^{(tot)}$ from the Landau level (2,0), particularly in the $B_\perp$ range where these peaks are resolved (see Fig. 3).

In the range of low $B_\perp$, where the resonances of individual transitions overlap and the dependence becomes weakly oscillatory, a decrease in the total rate $1/\tau_i^{(tot)}$ is observed when adding $B_\parallel$, despite an increase in the amplitude of the resonance peaks of individual transitions. The reason for this behavior can be seen in Figure 3b. The shift of resonances leads to the fact that in the range of $B_\perp$, in which the weakly oscillatory dependence takes place, there remain resonances of transitions with smaller $\Delta n$ and, as a consequence, with the lower amplitude $A_{LO}$. The summation of close resonances with lower amplitudes results in a lower value of $1/\tau_i^{(tot)}$.

Note that this behavior of the weakly oscillating part agrees with the limiting transition at $B_\perp \to 0$ – in the case of continuous 2D subbands, the lifetime at the bottom of the subband rises with increasing the inter-subband distance [5].

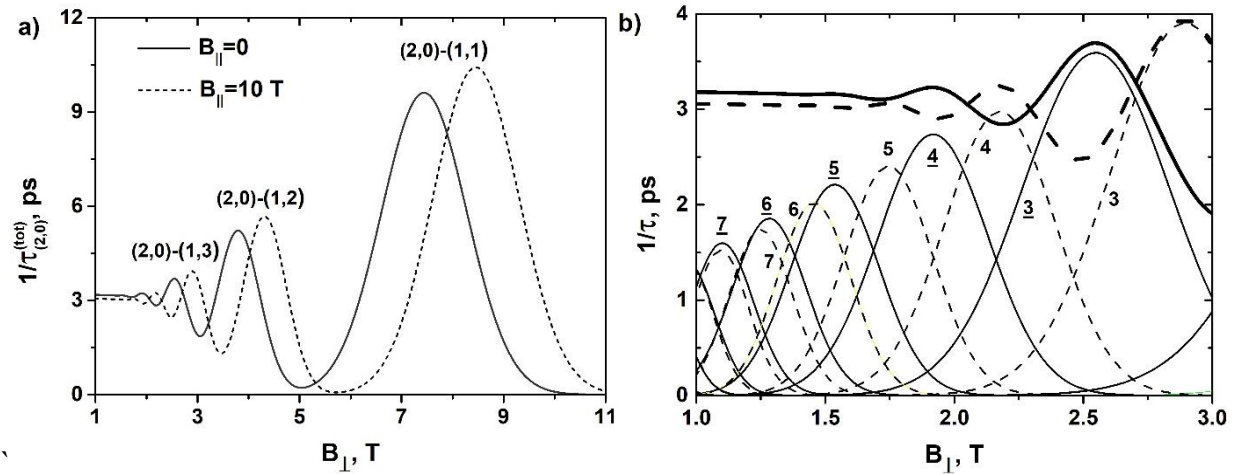

**Figure 3.** Total electron scattering rate from the (2,0) level due to optical phonon emission as a function of the quantizing magnetic field (solid curve) and a similar dependence when adding a magnetic field parallel to the quantum well layers (dashed curve) for a structure with an inter-subband distance exceeding the optical phonon energy. The curve marked by the number n is the rate of the individual transition $(2,0) \to (1,n)$. The data are given for a GaAs/Al$_{0.3}$Ga$_{0.7}$As quantum well of 15 nm width and $B_\parallel$=10 T.

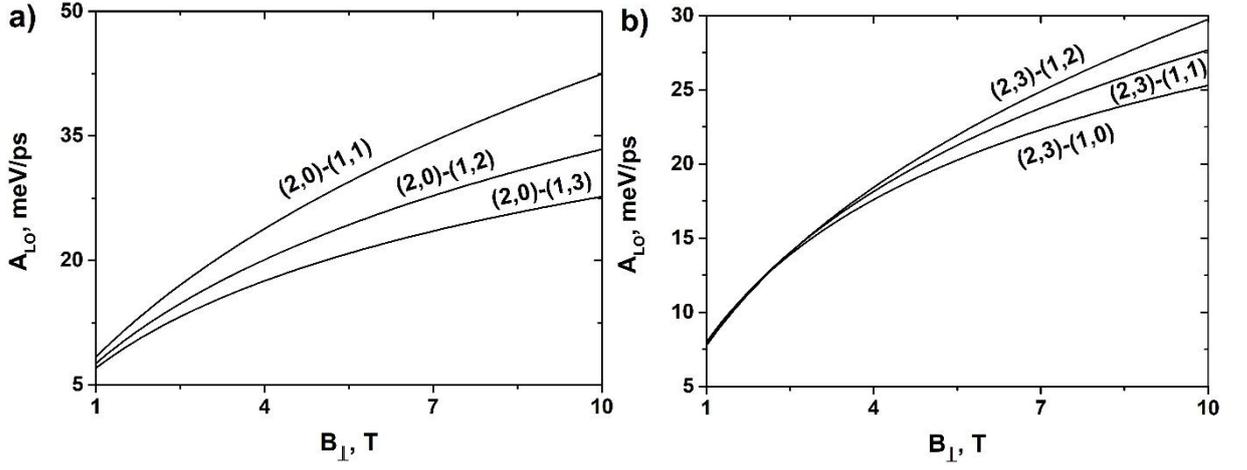

**Figure 4.** Dependence of the inter-subband transition amplitude on the quantizing component $B_\perp$ of the magnetic field: (a) $\Delta\varepsilon_{if} > \hbar\omega_{LO}$, well width $a$=15 nm; (b) $\Delta\varepsilon_{if} < \hbar\omega_{LO}$, $a$= 25 nm. Data are given for a GaAs/Al$_{0.3}$Ga$_{0.7}$As quantum well.

2). The dependence of the scattering rate on the magnetic field is principally different when the inter-subband distance is equal to the optical phonon energy ($\Delta\varepsilon_{if} = \hbar\omega_{LO}$). In this case, the resonance condition (30) at $B_\parallel = 0$ takes the form

$$n_i - n_f = 0. \tag{41}$$

Therefore, in this case, the selection rule «$\Delta n = 0$» arises - the energy conservation law is satisfied only for transitions between Landau levels with the same numbers. Moreover, for these transitions, the energy conservation law is satisfied at any value of the magnetic field. As a result, the scattering rate monotonically increases with increasing $B_\perp$ (Fig. 5), which differs significantly from the oscillatory dependence widely presented in the literature [7,8,26,27,28] in the case of $\Delta\varepsilon_{if} > \hbar\omega_{LO}$.

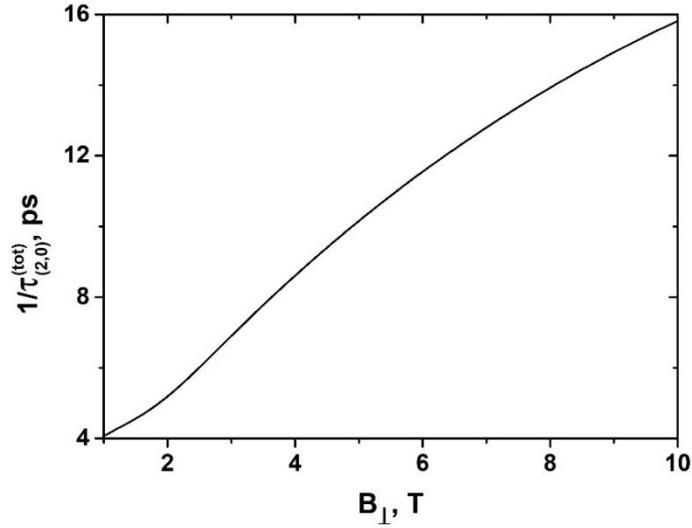

**Figure 5**: Magnetic field dependence of the total electron scattering rate from the Landau level (2,0) with optical phonon emission in a quantum well with an inter-subband distance equal to the optical phonon energy. The data are given for the GaAs/Al$_{0.3}$Ga$_{0.7}$As quantum well of 18.1 nm width.

In the tilted magnetic field ($B_\parallel \neq 0$), the resonance condition becomes

$$\delta\varepsilon_{if}(B_\parallel) + (n_i - n_f) \cdot \hbar\omega_\perp = 0. \qquad (42)$$

In this case, the relative shift of the subbands $\delta\varepsilon_{if}(B_\parallel)$ induced by the component $B_\parallel$, violates the resonance condition for the transitions with $n_i = n_f$, and, as a consequence, these transitions are suppressed (Fig. 6a). At the same time, an increase in the distance between the subbands $\delta\varepsilon_{if}(B_\parallel)$ makes it possible to fulfill the energy conservation law for transitions with $n_f > n_i$, and, as a consequence, to the appearance of intense transitions with $\Delta n > 0$ (Fig. 6b). As a result, as $B_\parallel$ increases, the monotonic dependence of the total scattering rate undergoes a qualitative change (Fig. 7), gradually transforming to the oscillatory dependence observed when the distance between the subbands exceeds the optical phonon energy.

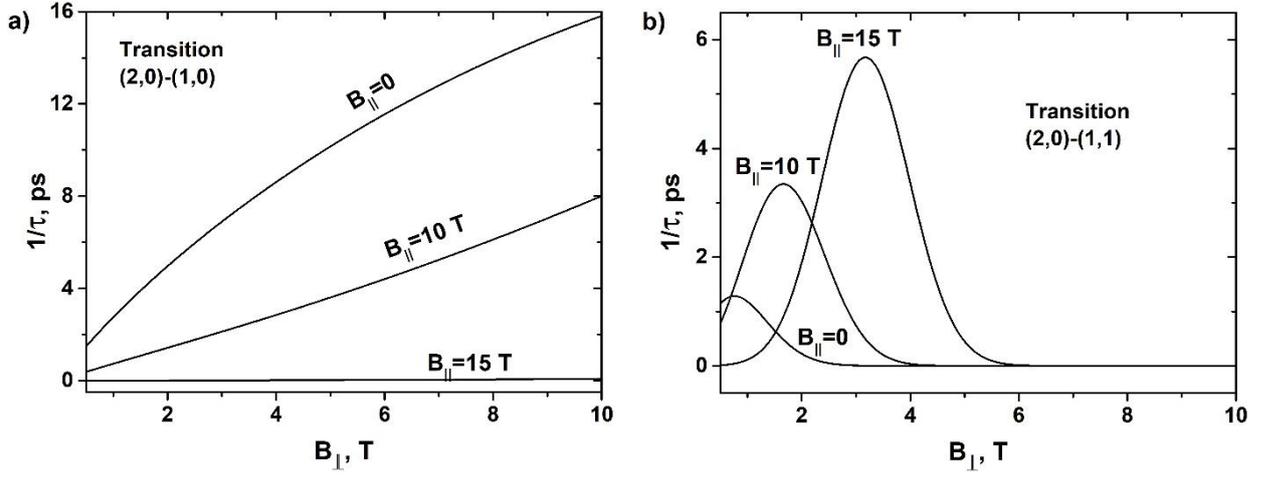

**Figure 6**: Dependence of the scattering rate with optical phonon emission on the magnetic field component $B_\perp$ at different values of $B_\parallel$, in the case when the distance between the subbands is equal to the optical phonon energy ($\Delta\varepsilon_{if} = \hbar\omega_{LO}$). a) transition $(2,0) \to (1,0)$; b) transition $(2,0) \to (1,1)$. The data are given for the GaAs/Al$_{0.3}$Ga$_{0.7}$As quantum well of 18.1 nm width.

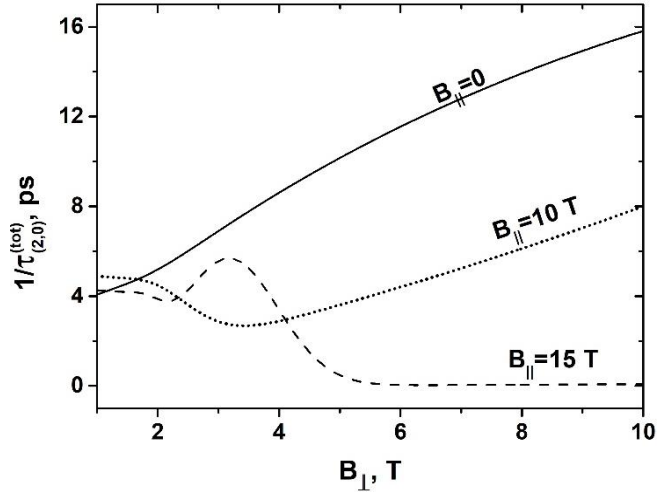

**Figure 7**: Dependence of the total scattering rate from the Landau level (2,0) on the quantizing component $B_\perp$ of the magnetic field at different values of $B_\parallel$ for the case when the distance between the subbands is equal to the optical phonon energy ($\Delta\varepsilon_{if} = \hbar\omega_{LO}$). The data are given for the GaAs/Al$_{0.3}$Ga$_{0.7}$As quantum well of 18.1 nm width.

We note that this curve transformation allows suppression of the scattering rate in the tilted magnetic field by an order of magnitude (Fig.8).

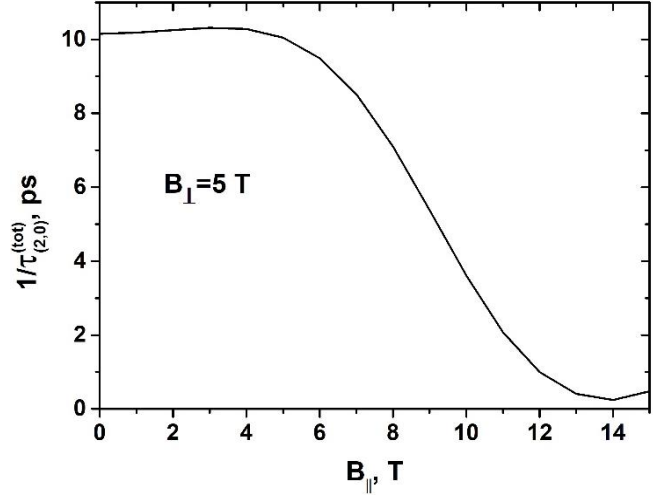

**Figure 8**. Dependence of the total scattering rate from the Landau level (2,0) on $B_\parallel$ at the fixed value of the quantizing component of the magnetic field $B_\perp = 5\,\text{T}$ in a GaAs/Al$_{0.3}$Ga$_{0.7}$As quantum well of 18.1 nm width.

In the case of $\Delta\varepsilon_{if} < \hbar\omega_{LO}$ the resonance does not occur for transitions with $\Delta n = 0$. Therefore, we can speak here of a selection rule «$\Delta n \neq 0$». Thus, at the transition from the case $\Delta\varepsilon_{if} < \hbar\omega_{LO}$ to the case $\Delta\varepsilon_{if} = \hbar\omega_{LO}$, the selection rule «$\Delta n \neq 0$» passes to the diametrically opposite selection rule «$\Delta n = 0$».

The transformation of the type of the scattering rate dependence on the quantizing magnetic field is caused by the $B_\parallel$- induced subband shift. However, the distance between the subbands can also be altered in other ways.

The most obvious of them is the variation of the quantum well width. Figure 9 shows the dependence of the scattering rate from the Landau level (2,0) on $B_\perp$ for several values of the GaAs/Al$_{0.3}$Ga$_{0.7}$As quantum well width ($B_\parallel = 0$). In such a quantum well, the condition $\Delta\varepsilon_{if} = \hbar\omega_{LO}$ is satisfied at a width of $a_0$=18.1 nm. The distance between the subbands decreases with increasing quantum well width. Therefore, at $a>a_0$, the dependence of the total scattering rate on $B_\perp$ remains monotonic, and with increasing $a$, scattering on phonons is gradually suppressed due to the finite transition broadening (Fig. 9a). With a decrease in the quantum well width below the value of $a_0$, a transformation of this dependence occurs, similar to that which we observed in a tilted magnetic field (Fig. 9b).

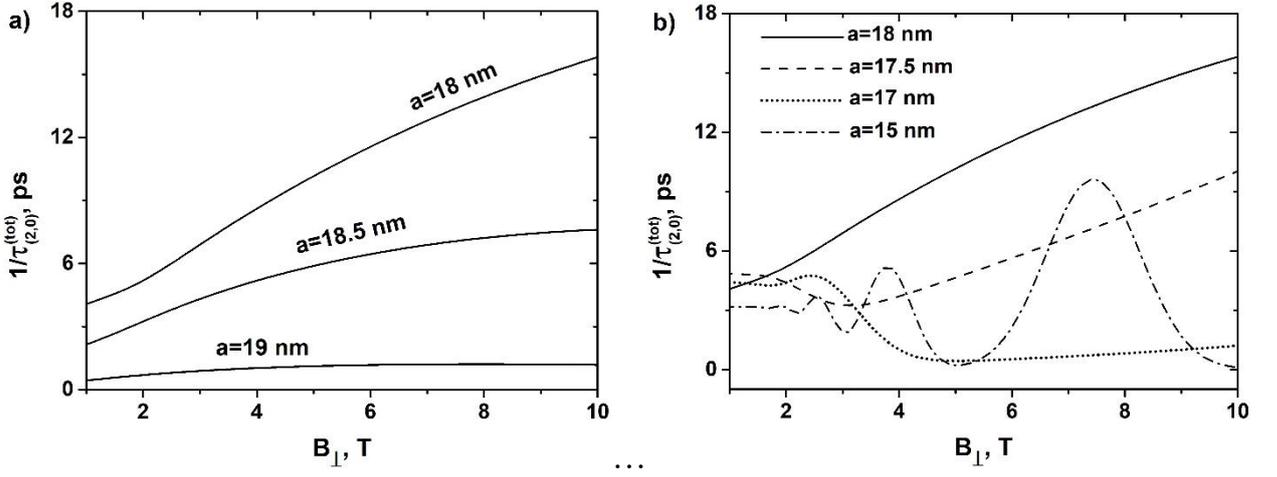

**Figure 9.** Dependence of the total scattering rate from the Landau level (2.0) on the magnetic field $B_\perp$ directed perpendicular to the layers, for different values of the GaAs/Al$_{0.3}$Ga$_{0.7}$As quantum well width $a$.

Another way to implement this effect is to apply a transverse electric field to the quantum well. The electric field increases the distance between two lower subbands (Fig. 10) and thus allows for the implementation of the indicated transformation of the scattering rate dependence on $B_\perp$ (Fig. 11). Note that by varying the electric field in a reasonable range, it is possible to change the scattering rate by an order of magnitude (Fig. 12).

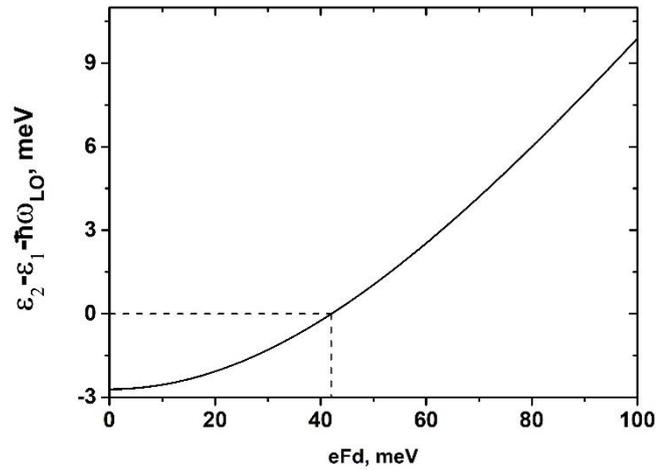

**Figure 10.** Dependence of the difference in the distance between the first and second subbands and the optical phonon energy on the voltage drop across the quantum well width. The data are given for a GaAs/Al$_{0.3}$Ga$_{0.7}$As quantum well with a width of 19 nm.

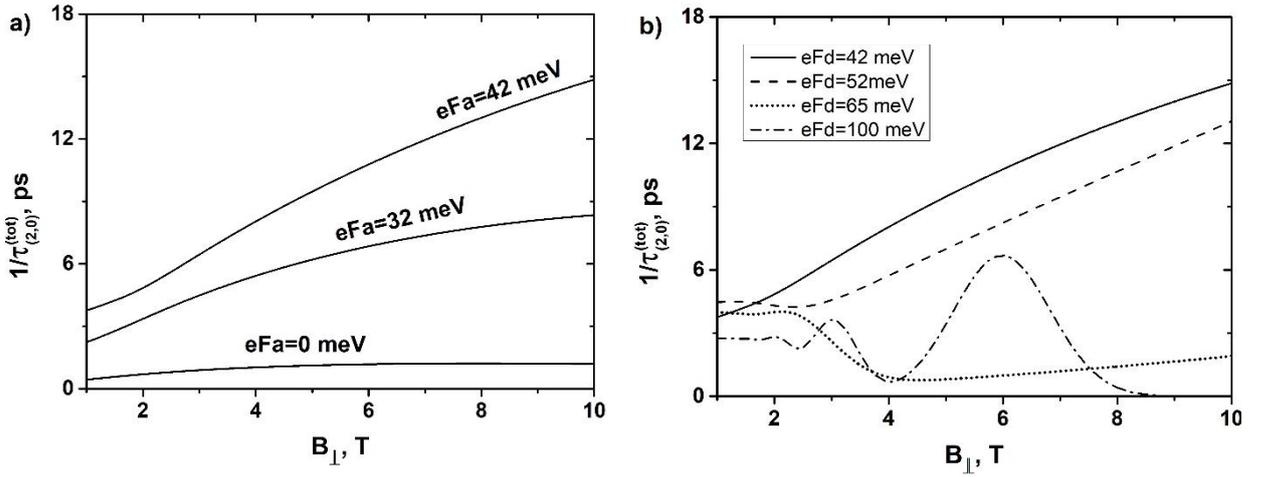

**Figure 11**. Dependence of the total scattering rate from the Landau level (2.0) on the magnetic field for different values of the voltage drop *eFa* across the GaAs/Al$_{0.3}$Ga$_{0.7}$As quantum well. The magnetic and electric fields are directed perpendicular to the quantum well layers. The quantum well width *a*=19 nm.

Variation of the quantum well width or application of the transverse electric field also leads to a change in the transition amplitude $A_{LO}$ (Fig. 13 and Fig. 14) due to a change in the subband wave functions $\varphi_\nu(z)$ – the amplitude increases with increasing quantum well width and decreases with increasing electric field strength. However, these changes are a rather weak and do not make an essential contribution to the effect.

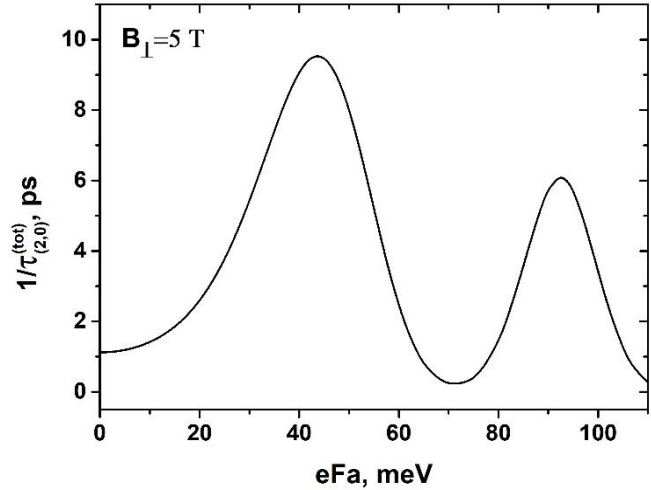

**Figure 12**. Dependence of the total scattering rate from the Landau level (2.0) on *eFa* at a fixed magnetic field value $B_\perp = 5\,\text{T}$ in a 19-nm wide GaAs/Al$_{0.3}$Ga$_{0.7}$As quantum well.

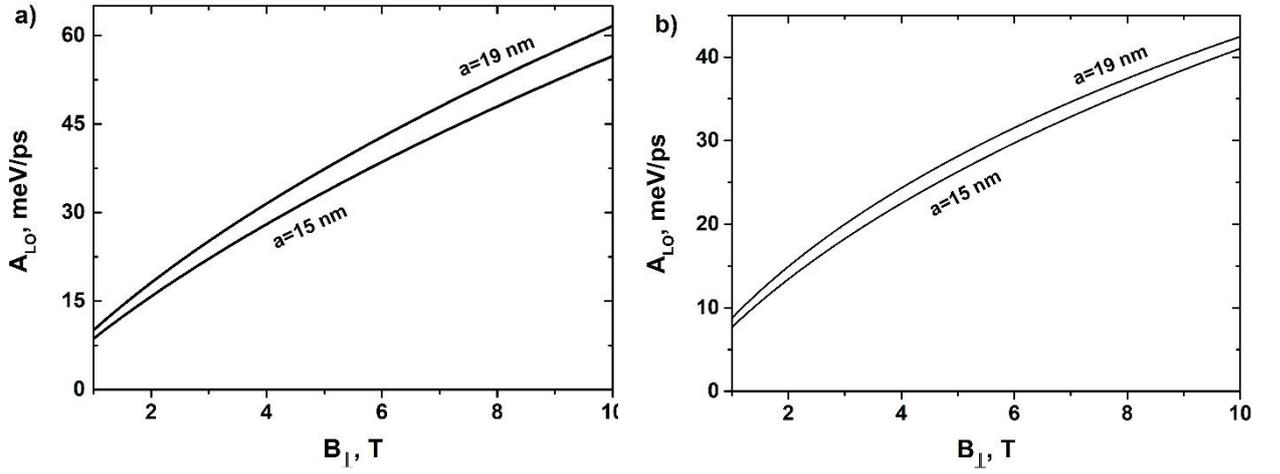

**Figure 13.** Dependence of the amplitude $A_{LO}$ of the transition $(2,0) \rightarrow (1,n)$ on the magnetic field component $B_\perp$ for different values of quantum well width $a$. a) n=0; b) n=1. The data are given for the GaAs/Al$_{0.3}$Ga$_{0.7}$As quantum well.

As we see, when the distance between the subbands is lower than the optical phonon energy ($\Delta\varepsilon_{if} < \hbar\omega_{LO}$), the scattering rate from the 0-th Landau level of the upper subband is suppressed when the resonance detuning $\Delta\varepsilon_{if} - \hbar\omega_{LO}$ is greater than the transition broadening.

However, in this case, transitions from higher Landau levels of the upper subband are of interest. Such transitions play a significant role in the relaxation of the energy of electrons located at Landau levels lying below the energy of the optical phonon [28-31]. Therefore, the question of the rate of intra-subband scattering processes and inter-subband scattering processes from the upper Landau levels is very important in this case.

As for the intra-subband optical phonon scattering processes, as has already been said, $B_\parallel$ does not affect them. The dependence of these processes on $B_\perp$ has a resonant character for transitions with $n_i > n_f$ (Fig. 15a). Also, one can note a rather weak decreasing dependence of the scattering rate in resonance on the width of the quantum well (Fig. 15b).

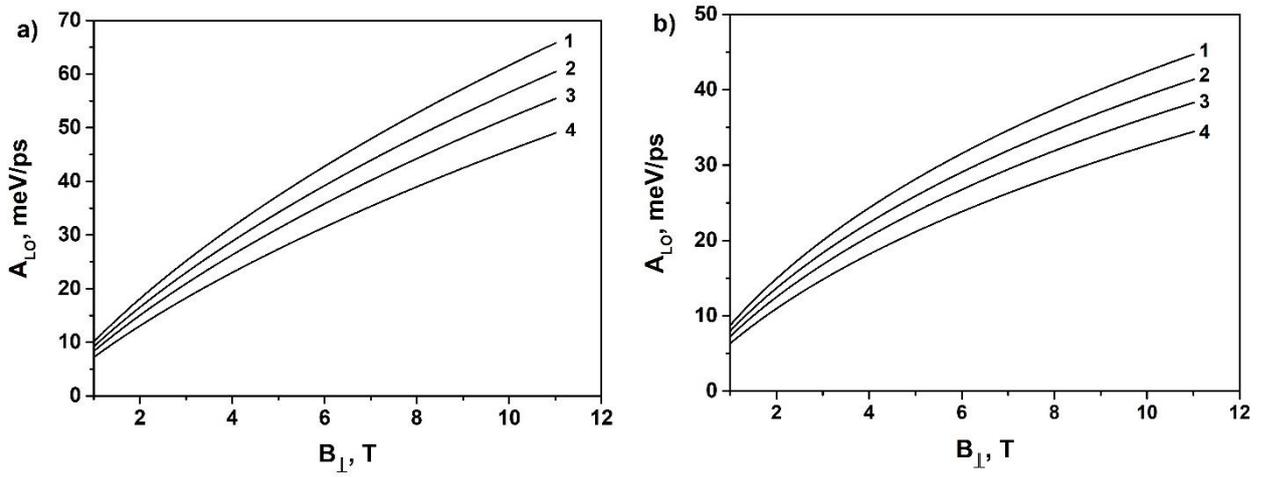

**Figure 14**. Dependence of the amplitude $A_{LO}$ of the transition $(2,0) \to (1,n)$ on the magnetic field $B_\perp$ for different values of the voltage drop $eFa$ across the width $a$ of the GaAs/Al$_{0.3}$Ga$_{0.7}$As quantum well. a) n=0; b) n=1. 1 – $eFa$=0; 2- $eFa$ =42 meV; 3 – $eFa$ =65 meV; 4 – $eFa$ =100 meV.

3). As can be seen from (34), if $\Delta\varepsilon_{if} < \hbar\omega_{LO}$ the resonances for intersubband transitions from the upper subband to the lower one occur only at $\Delta n < 0$. The application of a longitudinal magnetic field $B_\parallel$ leads to a shift in the resonance position to the value of $B_\perp$, determined by expression (35) (Fig. 16). However, in this case $\Delta n < 0$ and the shift occurs toward smaller values of $B_\perp$. The value of the scattering rate $1/\tau_{i \to f}$ in resonance decreases since the transition amplitude $A_{LO}$ decreases with lowering $B_\perp$.

The absolute value of the relative resonance shift increases proportionally to $B_\parallel^2$. Moreover, in accordance with the estimates made above (expression (38)), the proportionality coefficient increases significantly with the growth of the quantum well width $a$ (approximately proportionally to $a^2$). Therefore, in wide quantum wells with an inter-subband distance smaller than the optical phonon energy, the effect of $B_\parallel$ on the transition rates is quite significant (Fig. 17).

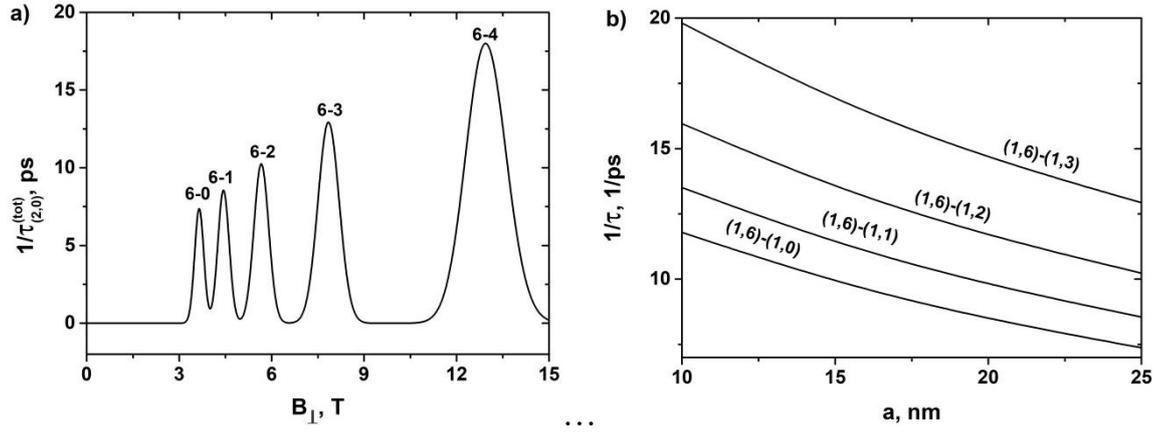

**Figure 15**. a) Dependence of the total rate of intra-subband scattering from the Landau level on the magnetic field. Resonances $(1, n_i) \to (1, n_f)$ are designated as $n_i \to n_f$. The data are given for the Landau level (1,6) in a 25-nm wide GaAs/Al$_{0.3}$Ga$_{0.7}$As quantum well. b) Dependence of the intra-subband scattering rate in resonance on the GaAs/Al$_{0.3}$Ga$_{0.7}$As quantum well width.

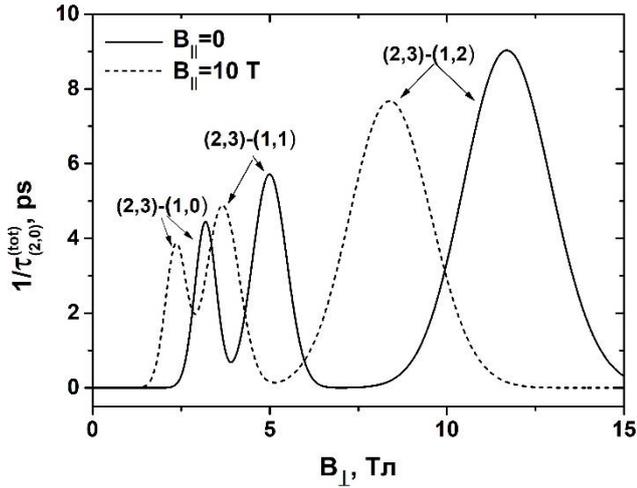

**Figure 16**. Total electron scattering rate from the Landau level (2,3) due to optical phonon emission as a function of the quantizing magnetic field $B_\perp$ (solid curve) and a similar dependence upon adding a magnetic field $B_\parallel$ parallel to the quantum well layers (dashed curve) for a structure where $\Delta\varepsilon_{if} < \hbar\omega_{LO}$. The data are given for a GaAs/Al$_{0.3}$Ga$_{0.7}$As quantum well of 25 nm width ($\Delta\varepsilon_{21} \approx 21$ meV $> \hbar\omega_{LO} \approx 36$ meV).

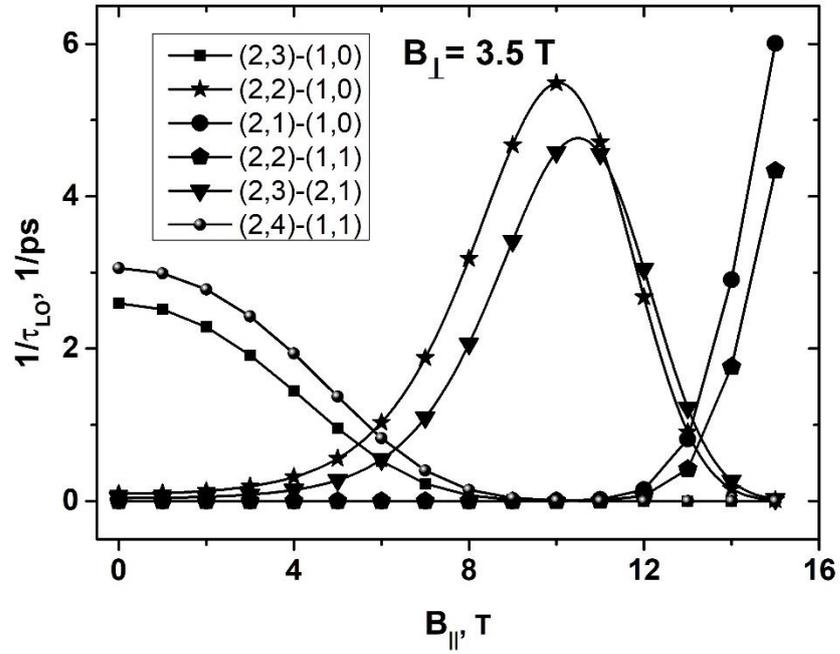

**Figure 17**. Dependence of the inter-subband scattering rate between different Landau levels of the first and second subbands on the parallel component of the magnetic field at a fixed value of quantizing component $B_\perp$ =3.5 T in a GaAs/Al$_{0.3}$Ga$_{0.7}$As quantum well of 25 nm width.

The above effects are mainly due to the energy conservation law in electron-phonon scattering and occur in quantum wells, including those with a symmetrical potential profile, in which the transition amplitudes $A_{LO}$ do not depend on $B_\parallel$.

In structures with an asymmetric potential profile, the subband wave-functions $\varphi(z)$ are no longer defined by parity, which implies that $\langle z \rangle_{v_i} \neq \langle z \rangle_{v_f}$. Consequently, the parameter $\xi_{v_i,v_f}$ becomes nonzero (see Fig. 18a). As a result, in these structures, the amplitude $A_{LO}$ of a transition becomes a function of $B_\parallel$ (Fig. 18b).

Finally, we note that all dependencies illustrated in the figures were calculated considering band nonparabolicity in an approach similar to that described in reference [38], utilizing second-order perturbation theory and accounting for terms in the expansion of the dispersion relation up to the fourth order in the wave vector. The nonparabolicity had small impact on the transition amplitudes, resulting only in a minor adjustment of the Landau levels and, consequently, the positions of the transition resonances.

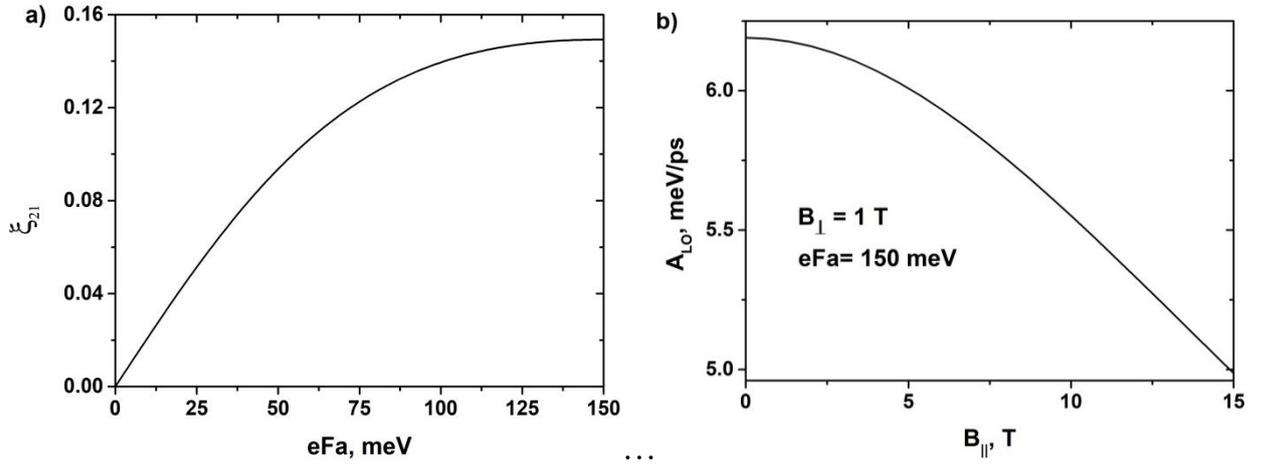

**Figure 18.** a) Parameter $\xi_{21}$ as a function of the electric field for transitions between two lower subbands of the GaAs/Al$_{0.3}$Ga$_{0.7}$As quantum well of 25 nm width. $B_\perp = B_\parallel = 1$ T; b) Dependence of the amplitude $A_{LO}$ of the transition $(2,0) \rightarrow (1,0)$ on $B_\parallel$ at a fixed quantizing component $B_\perp = 1$ T in a GaAs/Al$_{0.3}$Ga$_{0.7}$As quantum well of 25 nm width in a transverse electric field $F$. The voltage drop across the quantum well $eFa = 150$ meV.

## 5. Conclusion.

Expressions for the scattering rate of electrons with the emission of longitudinal polar optical phonons in a quantizing magnetic field tilted to the plane of the quantum well layers are derived. Through analytical integration, the calculation of the scattering rate for transitions between Landau levels is simplified to a double integral of a rapidly converging function.

It follows directly from the obtained expressions that in moderate magnetic fields (the cyclotron energy is less than the inter-subband distance), the rate of intra-subband scattering processes is determined by the magnetic field component $B_\perp$ that is perpendicular to the quantum well layers and is independent of the longitudinal component $B_\parallel$ of the magnetic field.

The effect of the longitudinal component $B_\parallel$ on the inter-subband scattering rate can be divided into two aspects.

The first aspect involves a change in electron energy when the magnetic field component $B_\parallel$ is applied. This component increases the distance between the subbands, leading to shifts in the resonances of inter-subband electron-phonon scattering to different values of the quantizing component $B_\perp$ of the magnetic field by a value $\delta B_\perp$ proportional to $B_\parallel^2$. The magnitude of this shift $\delta B_\perp$ is dependent on the width of the quantum well: the $|\delta B_\perp|$ increases with the width $a$ of the quantum well, approximately proportional to $a^2$.

The direction of the resonance shift is determined by the ratio of the optical phonon energy $\hbar\omega_{LO}$ to the distance between the subbands $\Delta\varepsilon_{if}$ when $B_\parallel = 0$. When $\Delta\varepsilon_{if} > \hbar\omega_{LO}$, the resonance moves toward higher values of $B_\perp$. Otherwise, when $\Delta\varepsilon_{if} < \hbar\omega_{LO}$, the resonances shift toward lower values of $B_\perp$.

An interesting situation is found when $\Delta\varepsilon_{if} = \hbar\omega_{LO}$. In this case, a selection rule $\Delta n = 0$ arises - resonances occur only for inter-subband transitions between the Landau levels with the same numbers. Moreover, the resonance condition holds for any value of the quantizing component $B_\perp$ of the magnetic field. As a result, the dependence of the scattering rate on $B_\perp$ is monotonic, which is drastically different from the widely known oscillating dependence at $\Delta\varepsilon_{if} > \hbar\omega_{LO}$. The transition from a monotonic dependence to an oscillating one can be achieved by additionally imposing a magnetic field $B_\parallel$ parallel to the layers of the structure, or a transverse electric field $F$. Such a transformation allows changing the lifetime of the Landau level by an order of magnitude by varying the $B_\parallel$ or $F$.

The second aspect of the effect of $B_\parallel$ on the electron-phonon scattering rate is the change in the electron wave function and, accordingly, the transition amplitude $A_{LO}$. The nature of the dependence of the inter-subband transition amplitude on $B_\parallel$ is essentially determined by the symmetry of the quantum well potential profile.

In symmetric quantum wells, he scattering amplitude of $A_{LO}$ is independent of $B_\parallel$, and the effect of the component $B_\parallel$ on the scattering rate consists of a shift of the resonances as a result of a subband shift caused by $B_\parallel$. With this shift, the maximum of the scattering rate $1/\tau_{i \to f}$ also varies due to the dependence of $A_{LO}$ on the quantizing component $B_\perp$ of the magnetic field. The transition amplitude $A_{LO}$ increases with increasing $B_\perp$. Consequently, when $\Delta\varepsilon_{if} > \hbar\omega_{LO}$ as the resonance shifts, there is an increase in the amplitude $A_{LO}$, leading to a rise in the maximum value of the scattering rate $1/\tau_{i\to f}$. Otherwise, when $\Delta\varepsilon_{if} > \hbar\omega_{LO}$, the resonance shift results in decrease of $1/\tau_{i\to f}$ in maximum.

In structures with an asymmetric potential profile, the amplitude depends on both $B_\perp$ and $B_\parallel$.